# Study of Monetization as a Way of Motivating Freemium Service Users

**Ilya V. Osipov**

Moscow Technological Institute, Moscow, Russia
Kedrova, 8/2, Moscow, 117292, Russia

**Alex A. Volinsky**

University of South Florida, Tampa, USA
4202 E. Fowler Ave., ENB118, Tampa FL 33620, USA

**Evgeny Nikulchev**

Moscow Technological Institute, Moscow, Russia
Kedrova, 8/2, Moscow, 117292, Russia
&
Higher School of Economics
Myasnitskaya str., 20, Moscow, 101000, Russia

**Dmitry Plokhov**

Moscow Technological Institute, Moscow, Russia
Kedrova, 8/2, Moscow, 117292, Russia

**Abstract**

The paper describes user behavior as a result of introducing monetization in the freemium educational online platform. Monetization resulted in alternative system growth mechanisms causing viral increase in the number of users. System metrics in terms of the *K-factor* was utilized as an indicator of the system user base growth. The weekly *K-factor* doubled as a result of monetization introduction.

**Keywords**: virality; retention; freemium; K-factor; metrics; open educational resource

# 1 Introduction

For the past 10 years the freemium model has become a prevailing strategy for developers of Internet services and mobile apps. The freemium concept implies a combined approach of providing services where customers can choose between using free version of the product with basic features or premium version for a fee. The premium status means that the product is enhanced by adding advanced features, such as new levels or characters in games, special working tools in programs and other application, or paying a fee rids users of watching mobile advertisement.

Typical functions of a freemium product aimed at the mass market are [1]:
(1) Basic main function of the application, the "core" cycle, for which the customers decide to use the application.
(2) Monetization function as an additional feature, which involves the most excited users.
(3) Retention is the cycle of users leaving the system with subsequent return.
(4) Viral function of the existing system users inviting new customers from the external environment.

There is a considerable number of research articles focused on different promotion methods of freemium products. Aral and Walker [2] examined viral promotion based on word-of-mouth peer influence through social network channels, including different features for communication. Another study [3] focused on identifying relations between satisfaction from technical features and emotional responses in users' willingness to share a social product with others. Bao and Chang [4] considered the influence of opinion leaders in social networks on product promotion and pointed at possibility of using disseminators in digital marketing. Approaches to selecting central users as influential seeds in the entire network are explored in the papers by Zhu [5], and Mochalova and Nanopoulos [6].

Given the specificity of the model, in the present work it is proposed to review popularity increase by means of a monetization mechanism in connection with the indicator of the system user base growth, which is also associated with another metrics – retention and virality. Evaluation of these metrics characterizing audience dynamics by studying a certain project is the subject of this article.

# 2 Experiment Defining

The online e-learning platform i2istudy, which grew to over 40,000 users, was used to conduct reported experiments and measure corresponding parameters [7, 8]. A basic idea of the i2istudy system is that instead of studying grammar, students have live interaction with foreigners to improve their oral communication skills. During the experiment demands of students learning Spanish were investigated. The developed platform uses direct audio-video live communication between users, along with the step-by-step teaching instructions combined with time tracking [8].

The developers used an idea of time banking [9] to track how much time is spent for teaching and learning foreign languages in the system. The program separately tracks teaching and learning time usage and provides features for selecting and connecting users. Time in minutes available to learn foreign languages from native speakers is a system internal currency. Figure 1 shows the system diagram before monetization, when users could only earn minutes by inviting friends and teaching other users. The scenario A is the referral generating process in which users can invite friends to expand the audience and receive 30 bonus minutes of system time for each invited registered user (gamification elements to motivate users to certain actions). The other scenarios B and C represent the core application cycle, where users spend most of the time teaching and learning from each other. In the reviewed experiments user behavior was affected by changing online system parameters to maximize the number of the system users. Although the i2istudy was envisioned as a free application, some monetization features were stipulated to positively influence user behavior.

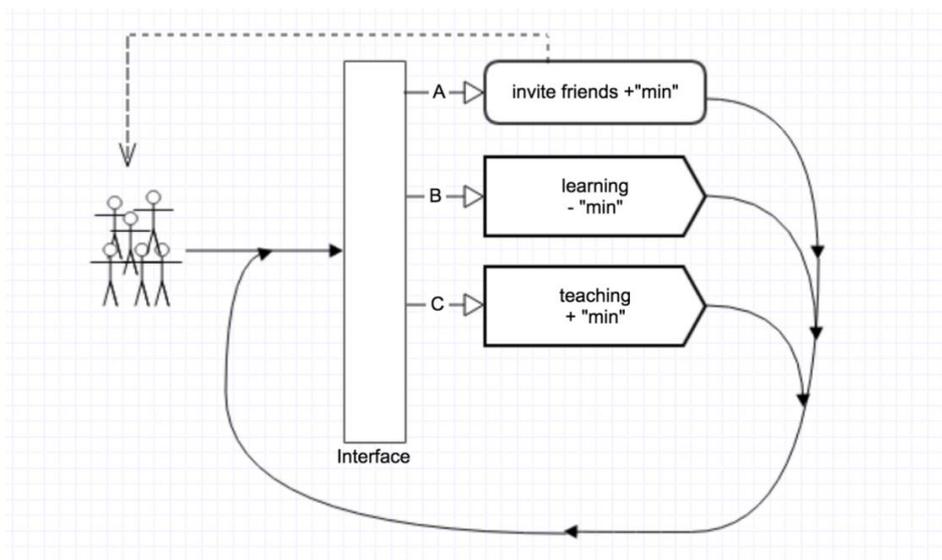

Figure 1. Diagram of the system state I, when users can only earn minutes by inviting friends and teaching other users.

As a result of the new feature implementation, the system was modified to the state II in Figure 2, which included an additional scenario D to purchase minutes. Different behavioral scenarios can either help or interfere with each other, depending on the situation [10].

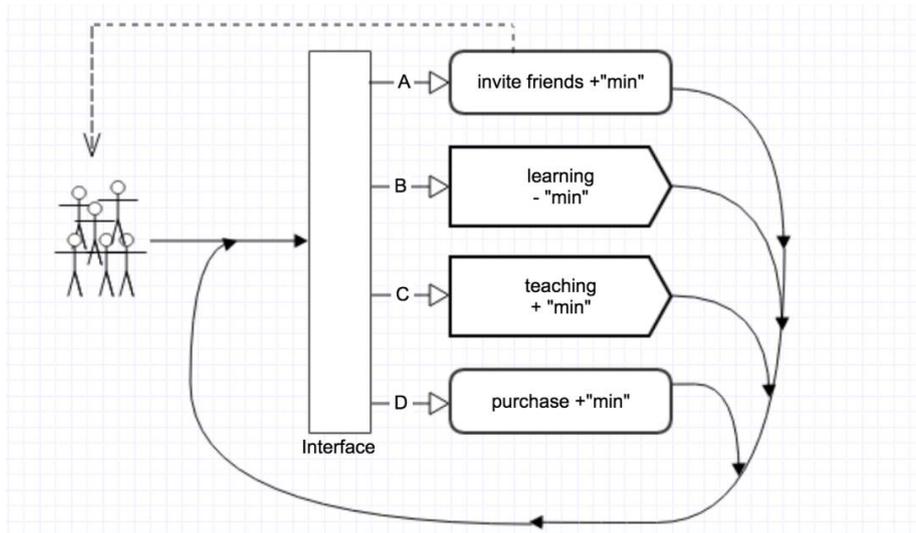

Figure 2. Diagram of the system state II after monetization was added.

## 3 Research Methods

Tracking of the metrics discussed above is the basis for monitoring audience growth and retention level for the freemium service. In the reviewed case these parameters are the audience growth, reflected by the viral *K-factor* coefficient, average user time spent in the system over a certain period (retention) and system monetization (percentage of paying users and how much money they spend).

Thus, the first objective is to grow the system in terms of the daily or weekly active users and virality in terms of the *K-factor*, which reflects the ratio of the virally attracted new users and active existing users over a certain time. The second objective is the system monetization at later development stages when there are enough users. However, it was decided to implement monetization during the system initial growth to study if the existing audience is ready to pay for the premium features.

To calculate the *K-factor*, one could use only the new users (*NU*), all users during a certain period (*U*), or only active users (*AU*). During the experiments only active users were considered to get more accurate results. Some sources use only novice users as the base [11], but in this study the total number of active users was included because they all contribute to virality, and not just the novice users. The weekly *K-factor* was calculated as:

$$K_{factor}^{weekly} = wIU/wAU \cdot 100\%, \qquad (1)$$

where *wIU* is the number of invited users and *wAU* is the number of active users in a given week.

Figure 3 shows the weekly *K-factor* dynamics with time. To make the *K-factor* calculations more objective, the raw data was corrected to exclude invitations sent by the system developers (Table 1).

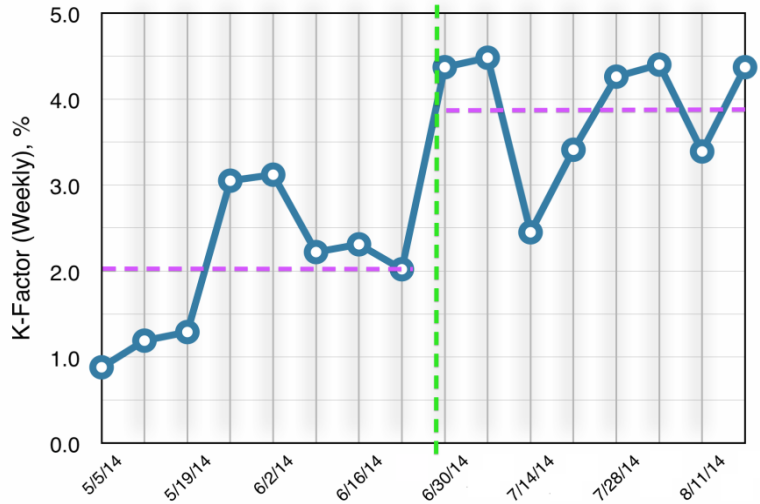

Figure 3. The weekly *K-factor* affected by the new monetization feature introduced on 6/28/2014.

**Table 1.** Data used for the weekly K-factor calculations in Figure 3.

| | 21.04-27.04 | 28.04-04.05 | 05.05-11.05 | 12.05-18.05 | 19.05-25.05 | 26.05-01.06 | 02.06-08.06 | 09.06-15.06 | 16.06-22.06 | 23.06-29.09 | 30.06-06.07 | 07.07-13.07 | 14.07-20.07 | 21.07-27.07 | 28.07-03.08 | 04.08-10.08 | 11.08-17.08 | 18.08-24.08 |
|---|---|---|---|---|---|---|---|---|---|---|---|---|---|---|---|---|---|---|
| Weekly Active Users *(wAU)* | | | 297 | 801 | 867 | 979 | 1080 | 1213 | 1827 | 2126 | 1763 | 2624 | 2572 | 1924 | 1716 | 1810 | 1576 | 947 |
| New Active Users *(wNU)* | | | 239 | 575 | 572 | 643 | 695 | 776 | 1358 | 1464 | 1140 | 1881 | 1643 | 1129 | 998 | 1066 | 824 | 358 |
| Invited Active Users *(wIU)* | | | 5 | 9 | 10 | 32 | 54 | 81 | 85 | 41 | 73 | 68 | 52 | 65 | 77 | 78 | 52 | 42 |
| K-Factor= wIU/wAU, % | | | 0.88 | 1.19 | 1.29 | 3.05 | 3.12 | 2.22 | 2.31 | 2.02 | 4.37 | 4.48 | 2.45 | 3.41 | 4.26 | 4.4 | 3.39 | 4.37 |

The *K-factor* over the whole system lifetime is often called the global *K-factor*, and calculated in terms of the conversion percentage, *IPi*, and the average number of invitations sent by each user, *AiPU*:

$$K_{factor} = AiPU \cdot IPi \qquad (2)$$

In turn, the *IPi* and *AiPU* values can be evaluated using a number of users who accepted the invitation, *IU*, the total number of sent invitations, *I*, and the total number of users, *U*:

$$IPi = IU/I \cdot 100\% \qquad (3)$$

$$AiPU = I/U \qquad (4)$$

## 4 Results and Discussion

To test the hypothesis and monetization concept, paid feature was added during the second month after the system launch. Attracting new users was marketed in the Facebook social network by placing ads in all four languages supported by the system (English, German, Spanish and Russian). As a result 40,000 users registered in the system along with over 1,000-1,500 daily site visits. However, since the paid feature had been implemented, only two $10 sales took place, which is by no means satisfactory for the freemium product. At the time of monetization implementation about 24,000 users were registered in the system at and 36% of them were involved in the process of teaching and learning.

At the same time, users started to send invitations more actively. The number of users willing to teach their native language rose from 48% to 55%, increasing user involvement in the teaching process. Thus, adding the option to purchase minutes indirectly improved system parameters of virality and retention, but had no effect on monetization.

To assess the system growth the weekly K-factor was used [12]. Mass mailing about the new paid feature caused a spike in the number of sent invitations, which affected the weekly K-factor in Table 1 and Figure 3. As seen in Figure 3, the weekly K-factor increased from 2.01±0.84% to 3.89±0.73% after the new feature announcement.

To test the hypothesis whether the K-factor changed as a result of monetization, statistical analysis was utilized. The null hypothesis was that the weekly K-factor did not change. The K-factor data before monetization were used as the expected values. The actual K-factor data for comparison were taken after monetization implementation. The calculated p-value was 0.01%, thus the hypothesis that the K-factor did not change as a result of monetization was rejected.

Therefore, in the observed case the monetization approach has not become effective for getting revenue. However, introduction of the paid feature has caused an increase in the system audience.

## 5 Conclusions

The initial attempt to monetize the e-learning platform by adding the paid feature affected the system in an unexpected way. Adding the option to purchase system minutes using real money did not result in significant sales, but motivated users to utilize the core cycle scenarios. It was expected that the users would either utilize or not utilize this option without changing the system performance parameters.

However, the paid option implementation resulted in the unexpected growth in viral mechanics of inviting friends. Thus, it was observed that direct motivation for a certain action within the system was ignored, but changed other indirect parameters (system growth in this case). It is quite clear and obvious that given alternative ways to earn virtual currency, users choose the most advantageous and simple ones for them. Viral system parameters improvement just due to the unrealized option to purchase minutes is an interesting effect observed in this study.

**References**

[1] F. Fields and B. Cotton, *Social Game Design: Monetization Methods and Mechanics*, Waltham, Morgan Kaufmann, 2012.

[2] S. Aral, and D. Walker, Creating Social Contagion through Viral Product Design: A Randomized Trial of Peer Influence in Networks, *Management Science*, **57** (2011), 1623 - 1639. http://dx.doi.org/10.1287/mnsc.1110.1421

[3] E. L. Cohen, What Makes Good Games Go Viral? The Role of Technology Use, Efficacy, Emotion and Enjoyment in Players' Decision to Share a Prosocial Digital Game, *Computers in Human Behavior*, **33** (2014), 321 - 329. http://dx.doi.org/10.1016/j.chb.2013.07.013

[4] T. Bao, and T.L.S. Chang, Finding Disseminators via Electronic Word of Mouth Message for Effective Marketing Communications, *Decision Support Systems*, **67** (2014), 21 - 29. http://dx.doi.org/10.1016/j.dss.2014.07.006

[5] Z. Zhu, Discovering the Influential Users Oriented to Viral Marketing Based on Online Social Networks, *Physica A: Statistical Mechanics and Its Applications*, **392** (2013), 3459 - 3469. http://dx.doi.org/10.1016/j.physa.2013.03.035

[6] A. Mochalova, and A. Nanopoulos, A Targeted Approach to Viral Marketing, *Electronic Commerce Research and Applications*, **13** (2014), 283 - 294. http://dx.doi.org/10.1016/j.elerap.2014.06.002

[7] I.V. Osipov, E. Nikulchev, A.A. Volinsky, A.Y. Prasikova, Study of Gamification Effectiveness in Online e-Learning Systems, *International Journal of Advanced Computer Science and Applications*, **6** (2015), 71 - 77. http://dx.doi.org/10.14569/ijacsa.2015.060211

[8] I.V. Osipov, A.Y. Prasikova, A.A. Volinsky, Participant Behavior and Content of the Online Foreign Languages Learning and Teaching Platform, *Computers in Human Behavior*, **50** (2015), 476 - 488. http://dx.doi.org/10.1016/j.chb.2015.04.028

[9] M. Marks, Time Banking Service Exchange Systems: A Review of the Research and Policy and Practice Implications in Support of Youth in Transition, *Children and Youth Services Review*, **34** (2012), 1230 - 1236. http://dx.doi.org/10.1016/j.childyouth.2012.02.017

[10] E.B. Seufert, *Freemium Economics: Leveraging Analytics and User Segmentation to Drive Revenue (The Savvy Manager's Guides)*. Waltham, Morgan Kaufmann, 2013.

[11] A.J. Kim, *Community Building on the Web*, Boston, Addison-Wesley Longman, 2000.

[12] F.F. Reichheld, The One Number You Need to Grow, *Harvard Business Review*, **81** (2003), 46 - 55.